\documentclass[12pt]{article}
\usepackage[left=2.5cm,top=2.50cm,right=2.5cm,bottom=2.50cm]{geometry}
\usepackage{mathrsfs}
\usepackage{amsmath,amssymb,latexsym,color,cancel,graphicx,bbm,colortbl}
\usepackage[english]{babel}
\usepackage[latin1]{inputenc}
\usepackage{ragged2e}
\begin{document}
\date{}

\title{Two-mode generalization of the Jaynes-Cummings and Anti-Jaynes-Cummings models}
\author{E. Chore$\tilde{n}$o$^{a}$\footnote{{\it E-mail address:} quike\_choreno@hotmail.com}, D. Ojeda-Guill\'en$^{b}$\\ M. Salazar-Ram\'irez$^{b}$ and V. D. Granados$^{a}$} \maketitle

\begin{minipage}{0.9\textwidth}
\small $^{a}$ Escuela Superior de F{\'i}sica y Matem\'aticas,
Instituto Polit\'ecnico Nacional, Ed. 9, Unidad Profesional Adolfo L\'opez Mateos, Delegaci\'on Gustavo A. Madero, C.P. 07738, Ciudad de M\'exico, Mexico.\\
\small $^{b}$ Escuela Superior de C\'omputo, Instituto Polit\'ecnico Nacional,
Av. Juan de Dios B\'atiz esq. Av. Miguel Oth\'on de Mendiz\'abal, Col. Lindavista,
Delegaci\'on Gustavo A. Madero, C.P. 07738, Ciudad de M\'exico, Mexico.\\

\end{minipage}

\begin{abstract}
We introduce two generalizations  of the Jaynes-Cummings (JC) model for two modes of oscillation. The first model is formed by two Jaynes-Cummings interactions, while the second model is written as a simultaneous Jaynes-Cummings and Anti-Jaynes-Cummings (AJC) interactions. We study some of its properties and obtain the energy spectrum and eigenfunctions of these models by using the tilting transformation and the Perelomov number coherent states of the two-dimensional harmonic oscillator. Moreover, as physical applications, we connect these new models with two important and novelty problems: The relativistic non-degenerate parametric amplifier and the relativistic problem of two coupled oscillators.
\end{abstract}

\section{Introduction}

The Dirac oscillator was introduced as an instance of a relativistic wave equation such that its non-relativistic limit leads to the well-known Schr\"odinger equation for the harmonic oscillator. This relativistic oscillator was introduced for the first time by Ito \emph{et al.} \cite{Ito} and Cook \cite{Cook}, and reintroduced later by Moshinsky and Szczepaniak \cite{Mos}. They added the linear term $-imc\omega\beta{\mathbf{\alpha}}\cdot \mathbf{r}$ to the relativistic momentum $\textbf{p}$ of the free-particle Dirac equation.

The Dirac-Moshinsky oscillator has been applied, among other problems, to quark confinement models in quantum chromodynamics, hexagonal lattices, and the emulation of graphene in electromagnetic billiards \cite{Sadurni1}. Moreover, the $1+1$ Dirac-Moshinsky oscillator has been mapped to the Anti-Jaynes-Cummings model by using the usual creation and annihilation operators \cite{Toyama,Nogami,Rados,Nos}. The exact solution of this problem was obtained by using the theory of the non-relativistic harmonic oscillator. Similarly, the $2+1$-dimensional Dirac-Moshinsky oscillator can be related to quantum optics via the Jaynes-Cummings and Anti-Jaynes-Cummings models by using the chiral creation and annihilation operators \cite{Bermudez,Chiral}.

The Jaynes-Cummings model \cite{Jaynes} plays an important role in quantum optics, since it describes the atom-field interaction. This is a fully quantized model and yet analytically solvable. It describes the interaction between a two-level atom and a quantized field. The exact solution of this model has been found by using the rotating wave approximation in the Hamiltonian of the two-level atom \cite{Haroche}. These solutions yield quantum collapse and revival of atomic inversion \cite{Narozhny}, squeezing of the radiation field \cite{Kuklinski}, among other quantum effects. All these effects have been corroborated experimentally, as can be seen in references \cite{Goy,Brune,Guerlin}. In recent years, there have been published some works on simulations of Dirac equation using different physical systems \cite{Lamata,Gerritsma,Lamata2}. These simulations allow to study relevant quantum relativistic effects, like the Zitterbewegung and Klein's paradox. In reference \cite{Johanning}, an extensive overview of current theoretical proposals and experiments for such quantum simulations with trapped ions is given.

The aim of the present work is to introduce two generalizations of the Jaynes-Cummings model with two modes of oscillation. We relate these new models with
the relativistic non-degenerate parametric amplifier and the relativistic problem of two coupled oscillators.

This work is organized as follows. In Section II, we review the main properties of $2+1$ Dirac oscillator and its exact mapping onto the Jaynes-Cummings model. In Sections III and IV, we introduce two generalizations of the Jaynes-Cummings model with two modes of oscillation. We use the Jordan-Schwinger realizations of the $su(1,1)$ and $su(2)$ Lie algebra to obtain the energy spectrum and eigenfunctions of these new JC-type models. We connect these models with the non-degenerate parametric amplifier and the problem of two coupled oscillators in the non-relativistic limit. Finally, we give some concluding remarks.

\section{The $2+1$ Dirac oscillator and Jaynes-Cummings model}

The time-independent Dirac equation for the Dirac-Moshinsky oscillator is given by the Hamiltonian
\begin{equation}
H_D\Psi=\left[c \mathbf{\alpha} \cdot\left(\mathbf{p}-im\omega \mathbf{r}\beta\right)+mc^2\beta\right]\Psi=E\Psi,\label{dirac}
\end{equation}
where $\alpha$ and $\beta$ are usual Dirac matrices, $m$ is the rest mass of the particle and $c$ is the speed of light. In the non-relativistic limit, this system reduces to a quantum simple harmonic oscillator (with mass $m$ and angular frequency $\omega$) with a strong spin-orbit coupling \cite{Mos}. The matrices $\alpha$ and $\beta$ must obey a Clifford algebra given by the anticommutation relations
\begin{equation}
\alpha_{j}\alpha_{k}+\alpha_{k}\alpha_{j}=2\delta_{jk}I, \quad\quad \alpha_{j}\beta+\beta\alpha_{j}=0.
\end{equation}
In $2+1$ dimensions, the solution to the Clifford algebra is given by the $2\times2$ Pauli matrices $\alpha_{x}=\sigma_{x}, \alpha_{y}=\sigma_{y}$ and  $\beta=\sigma_{z}$. In this case, the eigenfunction $|\Psi\rangle$ is described by a two-component spinor, where these components satisfy the coupled equations
\begin{equation}
\left(E-mc^2\right)|\Psi_1\rangle=c\left[(p_x+im\omega x)-i(p_{y}+im\omega y)\right]|\Psi_2\rangle,\label{c1}
\end{equation}
\begin{equation}
\left(E+mc^2\right)|\Psi_2\rangle=c\left[(p_x-im\omega x)-i(p_{y}-im\omega y)\right]|\Psi_1\rangle.\label{c2}
\end{equation}
The solutions of this problem can be found by introducing the chiral creation and annihilation operators (written in terms of the usual creation and annihilation operators $a_{x}, a_{x}^{\dag}$ and $a_{y}, a_{y}^{\dag}$)
\begin{equation}
a_{r}:=\frac{1}{\sqrt{2}}\left(a_{x}-ia_{y}\right), \quad\quad a_{r}^{\dag}:=\frac{1}{\sqrt{2}}\left(a_{x}^{\dag}+ia_{y}^{\dag}\right),
\end{equation}
\begin{equation}
a_{l}:=\frac{1}{\sqrt{2}}\left(a_{x}+ia_{y}\right), \quad\quad a_{l}^{\dag}:=\frac{1}{\sqrt{2}}\left(a_{x}^{\dag}-ia_{y}^{\dag}\right).
\end{equation}
These operators create a right or left quantum of angular momentum, respectively, and can be used to write the coupled equations (\ref{c1}) and (\ref{c2}) as \cite{Bermudez}
\begin{equation}
|\Psi_{1}\rangle=i\frac{2mc^{2}\sqrt{\xi}}{E-mc^{2}}a_{l}^{\dag}|\Psi_{2}\rangle\ ,  \quad\quad |\Psi_{2}\rangle=-i\frac{2mc^{2}\sqrt{\xi}}{E+mc^{2}}a_{l}|\Psi_{1}\rangle, \label{chiral}
\end{equation}
where the parameter $\xi=\hbar\omega/mc^{2}$ controls the non-relativistic limit.

The uncoupled equations for the two components $|\Psi_1\rangle$ and $|\Psi_2\rangle$ result to be
\begin{equation}
(E^{2}-m^{2}c^{4})|\Psi_{1}\rangle=4m^{2}c^{4}\xi a_{l}^{\dag}a_{l}|\Psi_{1}\rangle,\label{un2d1}
\end{equation}
\begin{equation}
(E^{2}-m^{2}c^{4})|\Psi_{2}\rangle=4m^{2}c^{4}\xi(1+a_{l}^{\dag}a_{l})|\Psi_{2}\rangle.\label{un2d2}
\end{equation}
Thus, the energy spectrum in terms of the quantum number $n_l$ is \cite{Bermudez}
\begin{equation}
E_{n_{l}}=\pm mc^{2}\sqrt{1+4\xi n_{l}}\label{2DE}.
\end{equation}

The $2+1$ Dirac-Moshinsky oscillator can be mapped onto the Jaynes-Cummings model of quantum optics as \cite{Bermudez}
\begin{align}
H&=2imc^{2}\sqrt{\xi}(a_{l}^{\dag}|\Psi_{2}\rangle\langle\Psi_{1}|-a_{l}|\Psi_{1}\rangle\langle\Psi_{2}|)+mc^{2}\sigma_{z}\nonumber\\&=\hbar(g\sigma_{-}a_{l}^{\dag}+g^{*}\sigma_{+}a_{l})+mc^{2}\sigma_{z},\label{H2+1}
\end{align}
where $g=2imc^{2}\sqrt{\xi}/\hbar$ is the coupling strength between orbital and spin degrees of freedom, and $\sigma_{\pm}$ are the spin raising and lowering operators. Formally, equation (\ref{H2+1}) is the Hamiltonian of the Jaynes-Cummings model of the quantum optics. Similarly, the $2+1$-dimensional Dirac-Moshinsky oscillator can be mapped onto the Anti-Jaynes-Cummings model by substituting $\omega\longrightarrow -\omega$ into equation (\ref{dirac}). This change leads to regard the excited state as the ground state, with a simultaneous change of sign in the detuning.

\section{The two-mode Jaynes-Cummings-Anti-Jaynes-Cummings model}

In order to connect the JC and AJC models with more general problems, we can introduce two modes of oscillation $a$ and $b$ to these models as follows
\begin{equation}
H_{JC}=\hbar(g\sigma_-a^{\dag}+g^*\sigma_+a)+mc^2\sigma_z\label{JC1},
\end{equation}
\begin{equation}
H_{AJC}=\hbar(f\sigma_-b+f^*\sigma_+b^{\dag})+mc^2\sigma_z\label{AJC}.
\end{equation}
With these two definitions we can propose a generalized JC-AJC model as a linear combination of equations (\ref{JC1}) and (\ref{AJC}) as follows
\begin{equation}
H=\hbar\left[\sigma_-(ga^{\dag}+f b)+\sigma_+(g^{*}a+f^*b^{\dag})\right]+mc^2\sigma_z,\label{GAJC}
\end{equation}
where $g$ and $f$ are two general complex parameters. This model will be connected later with a particular system by choosing suitably these complex parameters.

The coupled equations for the spinor components $|\Psi_1\rangle$ and $|\Psi_2\rangle$ are
\begin{equation}
\hbar(ga^{\dag}+fb)|\Psi_2\rangle=(E-mc^2)|\Psi_1\rangle,\label{1}
\end{equation}
\begin{equation}
\hbar(g^{*}a+f^*b^{\dag}|\Psi_1\rangle=(E+mc^2)|\Psi_2\rangle.\label{2}
\end{equation}
Uncoupling the above equations we found that the equations for $|\Psi_1\rangle$ and $|\Psi_2\rangle$ result to be
\begin{equation}
\hbar^2(|g|^2a^{\dag}a+|f|^2b^{\dag}b+fg^{*}ab+f^*ga^{\dag}b^{\dag}+|f|^2)|\Psi_1\rangle=(E^2-m^2c^4)|\Psi_1\rangle,\label{un1}
\end{equation}
\begin{equation}
\hbar^2(|g|^2a^{\dag}a+|f|^2b^{\dag}b+fg^{*}ab+f^*ga^{\dag}b^{\dag})|\Psi_2\rangle=(E^2-m^2c^4)|\Psi_2\rangle.\label{un2}
\end{equation}
Since these equations have the same mathematical structure, we will focus only on the equation for $|\Psi_1\rangle$. Then, the uncoupled equation
for $|\Psi_1\rangle$ can be written as
\begin{align}
\hbar^2&\left[\frac{1}{2}\left(|g|^2+|f|^2\right)(a^{\dag}a+b^{\dag}b+1)+fg^{*}ab+f^*ga^{\dag}b^{\dag}+\frac{1}{2}\left(|f|^2-|g|^2\right)(b^{\dag}b-a^{\dag}a+1)\right]|\Psi_1\rangle\nonumber\\ &=(E^2-m^2c^4)|\Psi_1\rangle.\label{unc1f}
\end{align}
Thus, by using the Jordan-Schwinger realization of the $su(1,1)$ algebra (see equation (\ref{gen11}) of Appendix), the above equation for $|\Psi_1\rangle$ becomes
\begin{align}
\hbar^2&\left[\left(|g|^2+|f|^2\right)K_{0}+fg^{*}K_{-}+f^*gK_{+}+\frac{1}{2}\left(|f|^2-|g|^2\right)(b^{\dag}b-a^{\dag}a+1)\right]|\Psi_1\rangle\nonumber\\ &=(E^2-m^2c^4)|\Psi_1\rangle. \label{unpsi}
\end{align}
The ladder operators $K_{\pm}$ can be removed in the Klein-Gordon-type Hamiltonian $H_{KG}|\Psi_{1}\rangle=\left(E^2-m^2  c^4\right)|\Psi_{1}\rangle$ by using the tilting transformation with the $SU(1,1)$ displacement operator $D(\xi)$. Therefore, if we apply the tilting transformation to both sides of equation (\ref{unpsi}) and we proceed as in references \cite{gerryberry,Nos1,Nos2}, we obtain
\begin{equation}
D^{\dag}(\xi)H_{KG}D(\xi)D^{\dag}(\xi)|\Psi_{1}\rangle =(E^{2}-m^{2}c^{4})D^{\dag}(\xi)|\Psi_{1}\rangle,
\end{equation}
\begin{equation*}
H'|\Psi_{1}'\rangle=(E^{2}-m^{2}c^{4})|\Psi_{1}'\rangle.
\end{equation*}
Notice that in these expressions $H'=D^{\dag}(\xi)H_{KG}D(\xi)$ is the tilted Hamiltonian and $|\Psi_{1}'\rangle$ its wave function. Thus, by using equations (\ref{DK+}), (\ref{DK-}) and (\ref{DK0}) of Appendix, the tilted Hamiltonian can be written as
\begin{align}
H'=& \hbar^{2}\left[\left((|f|^2+|g|^{2})(2\beta+1)+\frac{f^{*}g\xi^{*}\alpha}{|\xi|}+\frac{fg^{*}\xi\alpha}{|\xi|}\right)K_{0}\right.\nonumber\\&+ \left(\frac{(|f|^2+|g|^{2})\alpha\xi}{2|\xi|}+\frac{fg^{*}\beta\xi}{\xi^{*}}+ gf^{*}(\beta+1)\right)K_{+}\nonumber\\& + \left.\left(\frac{(|f|^2+|g|^{2})\alpha\xi^{*}}{2|\xi|}+fg^{*}(\beta+1)+\frac{f^{*}g\beta\xi^{*}}{\xi}\right)K_{-}\right]\nonumber\\&+\frac{\hbar^{2}}{2}(|f|^2-|g|^2)(b^{\dag}b-a^{\dag}a+1).\label{H2}
\end{align}
In this expression, the term $N_d+1=b^{\dag}b-a^{\dag}a+1$ remains unchanged under the tilting transformation, since it commutes with the $K_{\pm}$ operators (see Appendix). By choosing the coherent state parameters $\theta$ and $\varphi$ as
\begin{equation*}
\theta=\tanh^{-1}\left(\frac{2|f||g|}{|f|^2+|g|^2}\right),\quad\quad\varphi=i\ln{\left[\frac{(|f|^2+|g|^2)\alpha}{2fg^{*}(2\beta+1)} \right]},
\end{equation*}
the tilted Hamiltonian of equation (\ref{H2}) is reduced to
\begin{equation}
H'=\hbar^2\left(\frac{1}{2}(|f|^2-|g|^2)(N_d+1)+\sqrt{(|g|^{2}+|f|^2)^{2}-4|g|^{2}|f|^{2}}K_{0}\right).\label{hkg2}
\end{equation}

Let us now look the eigenfunctions $|\Psi_1'\rangle= D(\xi)^{\dag}|\Psi_1\rangle$ of $H'$. Since the operator $K_{0}$ is the Hamiltonian of the two-dimensional harmonic oscillator and commutes with $N_d+1$, we have that the eigenfunctions of $H'$ are given by
\begin{equation}
\psi'_{n_{l},m_n}(\rho,\phi)=\frac{1}{\sqrt{\pi}}e^{im_n\phi}(-1)^{n_{l}}\sqrt{\frac{2(n_{l})!}{(n_{l}+m_n)!}}\rho^{m_n}L^{m_n}_{n_{l}}(\rho^{2})e^{-1/2\rho^{2}}\label{Poly1},
\end{equation}
where $n_l$ is the left chiral quantum number.

From the action of the operators $a, a^{\dag}, b$ and $b^{\dag}$ on the basis $|n,m_n\rangle$, we have
\begin{equation}
K_{0}|n,m_n\rangle=\left(n_l+\frac{m_n}{2}+\frac{1}{2}\right)|n,m_n\rangle,\nonumber
\end{equation}
\begin{equation}
N_d|n,m_n\rangle=(b^{\dag}b-a^{\dag}a+1)|n,m_n\rangle=-(m_n-1)|n,m_n\rangle.
\end{equation}
In this $SU(1,1)$ representation, the group numbers $n,k$ are related with the physical numbers $n_l,m_n$ as $n=n_l$ and $k=\frac{1}{2}(m_n+1)$ \cite{Nos1}. Thus, from these results we can obtain that the energy spectrum of the generalized JC-AJC model is
\begin{equation}
E=\pm\sqrt{\hbar^2\left(\sqrt{(|g|^{2}+|f|^2)^{2}-4|g|^{2}|f|^{2}}\left(n_{l}+\frac{m_n}{2}+\frac{1}{2}\right)-\frac{1}{2}(|f|^2-|g|^2)(m_n-1)\right)+m^2c^4}.\label{spectrum2}
\end{equation}
Analogously, if we apply the same procedure to the uncoupled equation for the other spinor component $|\Psi_2\rangle$ we obtain
\begin{equation}
E=\pm\sqrt{\hbar^2\left(\sqrt{(|g|^{2}+|f|^2)^{2}-4|g|^{2}|f|^{2}}\left(n'_{l}+\frac{m'_n}{2}+\frac{1}{2}\right)-\frac{1}{2}(|f|^2-|g|^2)(m'_n+1)\right)+m^2c^4}.\
\end{equation}
Since both energies belong to the same solution, the energies must be the same. This leads to the fact that the spinor components $|\Psi_1\rangle$ and $|\Psi_2\rangle$ satisfy the relationships $n'_l\Rightarrow n_l+1$ and $m'_n\Rightarrow m_n-2$. It is worthwhile to mention that the energy spectrum (\ref{spectrum2}) can also be rewritten as
\begin{equation}
E_{n_{l}}=\pm{mc^{2}}\sqrt{1+\frac{2\hbar^2}{m^2c^4}(|f|^2-|g|^2)(n_{l}+1)}.\label{spectrums}
\end{equation}

The JC-AJC model of equation (\ref{GAJC}) can be connected with the standard $1+1$ Dirac-Moshinsky oscillator if we set $g=0$ and $f=\sqrt{\omega mc^2/\hbar}$. With this definition of the model parameters $f$ and $g$ the energy spectrum of equation (\ref{spectrums}) matches with those presented in references \cite{Nogami,Rados,Nos}.

The eigenfunctions of the generalized JC-AJC model are obtained from the relationship $|\Psi_1\rangle=D(\xi)|\Psi'_1\rangle$. Thus, by using the equations (\ref{Poly1}) and (\ref{PNCS}) of Appendix, we find that the action of the displacement operator $D(\xi)$ on $|\Psi'_1\rangle$ are the Perelomov number coherent states for the two-dimensional harmonic oscillator \cite{Nos1}
\begin{align*}
\Psi_{\zeta,k,n_{l}}^{(1)}=\langle\rho,\varphi|\zeta,k,n_{l}\rangle= &\frac{(-1)^{n_{l}}}{\sqrt{\pi}}e^{i(l-1/2)\phi}\sum_{s=0}^{\infty}\frac{\zeta^{s}}{s!}\sum_{j=0}^{n_{l}}\frac{(-\zeta^{*})^{j}}{j!}e^{\eta(k+n_{l}-j)}\frac{\sqrt{2\Gamma(n_{l}+1)\Gamma\left(n_{l}+l+\frac{1}{2}\right)}}{\Gamma\left(n_{l}-j+l+\frac{1}{2}\right)}\nonumber\\&\times\frac{\Gamma(n_{l}-j+s+1)}{\Gamma(n_{l}-j+1)}e^{-\rho^{2}/2}\rho^{l-1/2}L_{n_{l}-j+s}^{l-1/2}(\rho^{2}),\
\end{align*}
with $l=m_n+\frac{1}{2}$.
The above expression can also be rewritten as
\begin{align}
\psi_{n_{l},m_n}^{(1)}=&\sqrt{\frac{2\Gamma(n_{l}+1)}{\Gamma(n_{l}+m_n+1)}}\frac{(-1)^{n_{l}}}{\sqrt{\pi}}e^{im_n\phi}\frac{(-\zeta^{*})^{n_{l}}(1-|\zeta|^{2})^{\frac{m_n}{2}+\frac{1}{2}}(1+\sigma)^{n_{l}}}{(1-\zeta)^{m_n+1}}\nonumber\\&\times e^{-\frac{\rho^{2}(\zeta+1)}{2(1-\zeta)}}\rho^{m_n}L_{n_{l}}^{m_m}\left(\frac{\rho^{2}\sigma}{(1-\zeta)(1-\sigma)}\right),\label{eigen2}
\end{align}
where we have used $m_n=l-\frac{1}{2}$ and defined $\sigma$ as
\begin{equation*}
\sigma=\frac{1-|\zeta|^{2}}{(1-\zeta)(-\zeta^{*})}.
\end{equation*}
A similar result is obtained for the other spinor component $|\Psi_2\rangle$. Hence, we are able to construct the normalized spinor $|\Psi\rangle$ for the generalized JC-AJC model
\begin{equation}
|\Psi_{n_l,m_n}\rangle=\begin{pmatrix}
\sqrt{\frac{E\pm mc^2}{2E}}\psi_{n_l,m_n}^{(1)}\\
\mp i\sqrt{\frac{E\mp mc^2}{2E}}\psi_{n_l+1,m_n-2}^{(2)}
\end{pmatrix}.\label{estado2}
\end{equation}

Therefore, we have introduced a generalized JC-AJC model with two modes of oscillation. This new model was solved by using the tilting transformation and the $SU(1,1)$ Perelomov number coherent states for the two-dimensional harmonic oscillator.

\subsection{Special case: The relativistic non-degenerate parametric amplifier}

Now, we shall use the theory developed in Section III to give a connection between the generalized JC-AJC model and the relativistic non-degenerate parametric amplifier. Thus, for convenience we set the complex parameters $f$ and $g$ of the JC-AJC model as
\begin{equation}
g=i\sqrt{\frac{2mc^{2}\omega_{1}}{\hbar}}e^{-i\omega}, \quad\quad f=\sqrt{\frac{2mc^{2}\omega_{2}}{\hbar}} e^{i\omega}. \quad\quad\label{part2}
\end{equation}
If we substitute these parameters into the uncoupled equation (\ref{unc1f}) for the component $|\Psi_{1}\rangle$ we obtain
\begin{align}
H|\Psi_{1}\rangle &=2mc^{2}\hbar\left[\frac{1}{2}\left(\omega_{1}+\omega_{2}\right)(a^{\dag}a+b^{\dag}b+1)+2i\sqrt{\omega_{1}\omega_{2}} \left(a^{\dag}b^{\dag}e^{-2i\omega}-ab e^{2i\omega}\right)\right.\nonumber\\& + \left.\frac{1}{2}\left(\omega_{2}-\omega_{1}\right)(b^{\dag}b-a^{\dag}a+1)\right]|\Psi_{1}\rangle =(E^{2}-m^{2}c^{4})|\Psi_{1}\rangle. \label{HNONPA}
\end{align}
A similar equation holds for the component $|\Psi_{2}\rangle$. The tilted Hamiltonian of equation (\ref{hkg2}) is now given by
\begin{align}
H'|\Psi_{1}'\rangle &=\hbar mc^{2}\left[\left(\omega_{2}-\omega_{1}\right)(b^{\dag}b-a^{\dag}a+1)+2\sqrt{\left(\omega_{1}+\omega_{2}\right)^{2}-4\omega_{1}\omega_{2}}K_{0}\right]|\Psi_{1}'\rangle\nonumber\\&=(E^{2}-m^{2}c^{4})|\Psi_{1}'\rangle.
\end{align}
The energy spectrum of this problem for $f$ and $g$ defined as in equation (\ref{part2}) is obtained from the expression (\ref{spectrum2})
\begin{equation}
E=\pm\sqrt{\hbar mc^{2}\left(4\sqrt{\left(\omega_{1}+\omega_{2}\right)^{2}-4\omega_{1}\omega_{2}}\left(n_{l}+\frac{m}{2}+\frac{1}{2}\right)-\left(\omega_{2}-\omega_{1}\right)(m-1)\right)+m^{2}c^{4}},\label{energia}
\end{equation}
with the corresponding eigenfunctions given by equations (\ref{eigen2}) and (\ref{estado2}).

If we write the energy as $E=mc^{2}+\varepsilon$ in (\ref{HNONPA}), we obtain that in the non-relativistic limit $(\epsilon\ll mc^2)$ the uncoupled equation for $|\Psi_{1}\rangle$ becomes
\begin{align}
H|\Psi_{1}\rangle=\left[\hbar\omega_{1}a^{\dag}a+\hbar\omega_{2}b^{\dag}b+i\hbar\sqrt{\omega_{1}\omega_{2}}(a^{\dag}b^{\dag}e^{-2i\omega}-abe^{2i\omega})\right]|\Psi_{1}\rangle =\varepsilon|\Psi_{1}\rangle.
\end{align}
If we identify the coupling constant $\chi$ with the term $\sqrt{\omega_1\omega_2}$, we obtain the time-independent Hamiltonian of the non-degenerate parametric amplifier \cite{walls}.
Therefore, if we set the complex parameters $f$ and $g$ as those given by equation (\ref{part2}), our generalized JC-AJC model is reduced, in the non-relativistic limit $(\epsilon\ll mc^2)$, to the time-independent Hamiltonian of the non-degenerate parametric amplifier. Moreover, the energy spectrum of equation (\ref{energia}), in the non-relativistic limit, is in full agreement with the energy spectrum of the non-degenerate parametric amplifier \cite{Nos1}. Also, it is important to note that the definition of the parameters $f$ and $g$ of equation (\ref{part2}) is not unique, since they can be introduced in other form and the connection holds.

\section{The two-mode Jaynes-Cummings model}

In a similar way to the model proposed in Section III, we can consider two Jaynes-Cummings models $H_{JC_{1}}$ and $H_{JC_{2}}$ with two different modes of oscillation
\begin{equation}
H_{JC_{1}}=\hbar(g\sigma_{-}a^{\dag}+g^*\sigma_{+}a)+mc^2\sigma_z\label{JCa},
\end{equation}
and
\begin{equation}
H_{JC_{2}}=\hbar(f\sigma_{-}b^{\dag}+f^*\sigma_{+}b)+mc^2\sigma_z\label{JCb}.
\end{equation}
Therefore, we can propose a generalization of the Jaynes-Cummings model as follows
\begin{equation}
H=\hbar\left[\sigma_-(ga^{\dag}+f b^{\dag})+\sigma_+(g^{*}a+f^*b)\right]+mc^2\sigma_z,\label{GJC}
\end{equation}
where $g$ and $f$ are again two general complex parameters.

The coupled equations for the
spinor components $|\Psi_1\rangle$ and $|\Psi_2\rangle$ are
\begin{equation}
\hbar(ga^{\dag}+fb^{\dag})|\Psi_2\rangle=(E-mc^2)|\Psi_1\rangle,\label{nc1}
\end{equation}
\begin{equation}
\hbar(g^{*}a+f^{*}b|\Psi_1\rangle=(E+mc^2)|\Psi_2\rangle.\label{nc2}
\end{equation}

Hence, the uncoupled equations for the spinor components are easily obtained from above
expressions, which result to be
\begin{equation}
\hbar^2(|g|^{2}aa^{\dag}+|f|^{2}bb^{\dag}+g^{*}f ab^{\dag}+f^{*}gba^{\dag})|\Psi_2\rangle=(E^2-m^2c^4)|\Psi_2\rangle,\label{nu1}
\end{equation}
\begin{equation}
\hbar^2(|g|^{2}a^{\dag}a+|f|^{2}b^{\dag}b+fg^{*}b^{\dag}a+f^{*}ga^{\dag}b)|\Psi_1\rangle=(E^2-m^2c^4)|\Psi_1\rangle.\label{nu2}
\end{equation}
In this manner, by using the Jordan-Schwinger realization of the $su(2)$ Lie algebra (see equation (\ref{gen}) of Appendix),
in addition to the number operator $N_s=a^{\dag}a+b^{\dag}b$, we can write the uncoupled equation for $|\Psi_1\rangle$ as
\begin{equation}
\hbar^2\left[(|g|^2-|f|^2)J_{0}+f^{*}gJ_{+}+fg^{*}J_{-}+\frac{1}{2}(|f|^2+|g|^2)N_s\right]|\Psi_1\rangle=(E^2-m^2c^4)|\Psi_1\rangle\label{unc1}.
\end{equation}
Now, we can apply the tilting transformation to the above eigenvalue equation, in order to remove the ladder operators $J_{\pm}$
\begin{eqnarray}
\hbar^2D^{\dagger}(\xi)\left[(|g|^2-|f|^2)J_{0}+f^{*}gJ_{+}+fg^{*}J_{-}+\frac{1}{2}(|f|^2+|g|^2)N_s\right]D(\xi)D^{\dagger}(\xi)|\Psi_1\rangle\nonumber\\=(E^2-m^2c^4)D^{\dagger}(\xi)|\Psi_1\rangle,
\end{eqnarray}
where $D(\xi)$ is the $SU(2)$ displacement operator and $\xi=-\frac{1}{2}\theta e^{-i\varphi}$ (see Appendix). If we define the tilted Hamiltonian $H'=D^{\dagger}(\xi)HD(\xi)$ and the wave function $|\Psi'_1\rangle=D^{\dagger}(\xi)|\Psi_1\rangle$, this equation can be written as $H'|\Psi'_1\rangle=(E^2-m^2c^4)|\Psi'_1\rangle$. Moreover, since the operator $N_s$ commutes whit $J_+$ and $J_-$, it remains unchanged
under the tilting transformation. Therefore, we find that the tilted Hamiltonian results to be
\begin{align}
 H'=&\hbar^{2}\left[  \left((|g|^{2}-|f|^2)\left(2\epsilon + 1 \right) - \frac{f^{*}g \xi^{*} \delta}{|\xi|} - \frac{fg^{*}\xi \delta}{|\xi|}\right)J_{0}\right.\nonumber\\ & + \left(\frac{(|g|^{2}-|f|^{2})\delta\xi^{*}} {2|\xi|} + \frac{f^{*}g\epsilon \xi^{*}}{\xi}+ fg^{*}( \epsilon +1) \right)J_{-}\nonumber\\& + \left. \left( \frac{(|g|^{2}-|f|^{2})\delta\xi}{2|\xi|} + f^{*}g(\epsilon +1) +\frac{fg^{*}\epsilon \xi}{\xi^{*}}  \right)J_{+} \right]\nonumber\\&+\frac{\hbar^{2}}{2}(|f|^2+|g|^2)N_s\label{tildedH}.
\end{align}

By choosing the coherent state parameters $\theta$ and $\varphi$ as
\begin{equation}
\theta=\arctan\left(\frac{2|f||g|}{|f|^2-|g|^2}\right), \quad\quad \varphi=i\ln{\left[\frac{(|f|^2-|g|^2)\delta}{2fg(2\epsilon+1)} \right]},
\end{equation}
the tilted Hamiltonian $H'$ (\ref{tildedH}) is now written as
\begin{equation}
H'=|\Psi'_1\rangle=\hbar^2\left(\frac{1}{2}(|f|^2+|g|^2)N_s+\sqrt{(|g|^{2}-|f|^2)^{2}+4|g|^{2}|f|^{2}}J_{0}\right)|\Psi'_1\rangle=(E^2-m^2c^4)|\Psi'_1\rangle.\label{hkg}
\end{equation}
Since the operators $J_{0}$ and $N_s$ commutes, they share common eigenfunctions. Thus, by using that $N_s$ is the Hamiltonian of the two-dimensional harmonic oscillator, the eigenfunctions of the tilted Hamiltonian $H'$ are
\begin{equation}
\psi'_{n_{l},m_n}(\rho,\phi)=\frac{1}{\sqrt{\pi}}e^{im_n\phi}(-1)^{n_{l}}\sqrt{\frac{2(n_{l})!}{(n_{l}+m_n)!}}\rho^{m_n}L^{m_n}_{n_{l}}(\rho^{2})e^{-1/2\rho^{2}}\label{Poly}.
\end{equation}
From the action of the operators $a, a^{\dag}, b$ and $b^{\dag}$ on the basis $|n,m_n\rangle$, we have
\begin{equation}
J_{0}|n,m_n\rangle=\frac{m_n}{2}|n,m_n\rangle,\nonumber
\end{equation}
\begin{equation}
N_s|n,m_n\rangle=(b^{\dag}b+a^{\dag}a)|n,m_n\rangle=(2n_l+m_n)|n,m_n\rangle.
\end{equation}
In this $SU(2)$ representation, the group numbers $j,\mu$ are related with the physical numbers $n_l,m_n$ as $j=n_l+\frac{m_n}{2}$ and $\mu=\frac{m_n}{2}$ \cite{Nos2}. Thus, the energy spectrum of the generalized Jaynes-Cummings model is
\begin{equation}
E=\pm\sqrt{\hbar^2\left((|f|^2+|g|^2)\left(n_l+\frac{m_n}{2}\right)\pm\frac{1}{2}\sqrt{(|g|^{2}-|f|^2)^{2}+4|g|^{2}|f|^{2}}m_{n}\right)+m^2c^4}.\label{spectrum}
\end{equation}
If we apply the same procedure to the uncoupled equation for the other spinor component $|\Psi_2\rangle$ we obtain the energy spectrum
\begin{equation}
E=\pm\sqrt{\hbar^2\left((|f|^2+|g|^2)(n'_l+\frac{m'_n}{2}+1)\pm\frac{1}{2}\sqrt{(|g|^{2}-|f|^2)^{2}+4|g|^{2}|f|^{2}}m'_{n}\right)+m^2c^4}.\label{spec}
\end{equation}
Since both energies belong to the same eigenstate, this leads to take $n'_{l}\Rightarrow n_{l}-1$ in the spinor component $|\Psi_2\rangle$.

The energy spectrum of equation (\ref{spectrum}) can also be written as
\begin{equation}
E=\pm{mc^{2}}\sqrt{\frac{\hbar^2}{2m^{2}c^{4}}\left(|f|^2+|g|^2\right)\left(N\pm{m_n}\right)+1},\label{spectrumRE}
\end{equation}
where $N$ is the eigenvalue of the operator $N_s$. It is important to note that the JC-JC model of equation (\ref{GJC}) can be connected with the $1+1$ Dirac-Moshinsky oscillator if we set the model parameters as $g=0$ and $f=\sqrt{2mc^2\omega/\hbar}$. Similarly, the connection with the $2+1$ Dirac-Moshinsky oscillator can be established if we set $f=g=\sqrt{2mc^2\omega/\hbar}$.

The eigenfunctions of the generalized Jaynes-Cummings model are obtained from the relationship $|\Psi_1\rangle=D(\xi)|\Psi'_1\rangle$. Thus, by using equation (\ref{Poly}) we find that
the action of the displacement operator $D(\xi)$ on $|\Psi'_1\rangle$ is
\begin{align}
\Psi_{n_{l},m_{n},\zeta}^{(1)}=\langle\rho,\varphi|\zeta,n_{l},m_{n}\rangle= &\frac{e^{-\frac{\rho^{2}}{2}}}{\sqrt{\pi}}\sum_{s=0}^{n_{l}+n}\frac{\zeta^{s}}{s!}\sum_{n=0}^{n_{l}+m_{n}}\frac{(-\zeta^{*})^{n}}{n!}e^{\frac{\eta}{2}(m_{n}-2n)}e^{i(m_{n}-2n+2s)\varphi}(-1)^{n_{l}+n-s}\nonumber\\&\times\frac{\Gamma(n_{l}+n+1)}{\Gamma(n_{l}+m_{n}-n+1)}\left[\frac{2\Gamma(n_{l}+m_{n}+1)}{\Gamma(n_{l}+1)}\right]^{1/2}\rho^{(m_{n}-2n+2s)}L_{n_{l}+n-s}^{(m_{n}-2n+2s)}(\rho^{2}).\
\end{align}
A similar result holds for the other spinor component $\Psi_{n_{l},m_{n},\zeta}^{(2)}$. Thus, we are able to construct the normalized spinor $|\Psi\rangle$ in the form

\begin{equation}
|\Psi_{n_{l},m_{n},\zeta}\rangle=\begin{pmatrix}
\sqrt{\frac{E\pm mc^2}{2E}}\Psi_{n_{l},m_{n},\zeta}^{(1)}\\
\mp i\sqrt{\frac{E\mp mc^2}{2E}}\Psi_{n_{l}-1,m_{n},\zeta}^{(2)}
\end{pmatrix},\label{estado}
\end{equation}

In this way, we have introduced a generalized JC-JC model with two modes of oscillation. As in the case of the JC-AJC model, the solution was obtained by using the tilting transformation and the $SU(2)$ Perelomov number coherent states for the two-dimensional harmonic oscillator.

\subsection{Special case: The relativistic problem of two coupled oscillators}

In this section, we shall give a physical application of the theory developed in Section IV related to the two-mode generalization of the Jaynes-Cummings model. In particular, we shall give the connection between this generalized Jaynes-Cummings model and the relativistic problem of two coupled oscillators.

A particular case of the Hamiltonian proposed in equation (\ref{GJC}) is obtained if we set the complex parameters $g$ and $f$ as
\begin{equation}
g=\sqrt{\frac{2mc^{2}\omega_{1}}{\hbar}}e^{-i\phi}, \quad\quad f=\sqrt{\frac{2mc^{2}\omega_{2}}{\hbar}}e^{i\phi}. \quad\quad\label{part}
\end{equation}
Thus, the uncoupled equation for the upper component $|\Psi_1\rangle$ takes the form (see equation (\ref{nu2}))
\begin{equation}
H|\Psi_1\rangle=(E^2-m^2c^4)|\Psi_1\rangle, \label{pnu1}
\end{equation}
where the Hamiltonian written in terms of the Jordan-Schwinger realization of the $su(2)$ Lie algebra of equation (\ref{gen}) is
\begin{equation}
H=\hbar\left[2mc^{2}\left(\omega_{1}-\omega_{2}\right)J_{0}+2mc^{2}\sqrt{\omega_{1}\omega_{2}}(J_++J_-)+mc^{2}\left(\omega_{1}+\omega_{2}\right)N_s\right]].
\end{equation}
A similar equation holds for the lower component $|\Psi_2\rangle$. The transformations performed to the general
case (equations (\ref{unc1})-(\ref{hkg})), lead us to obtain the following tilted Hamiltonian $H'$
\begin{equation}
H'|\Psi'_1\rangle=2mc^{2}\hbar\left[\left(\omega_{1}+\omega_{2}\right)N_s+\sqrt{\left(\omega_{1}-\omega_{2}\right)^{2}+4\omega_{1}\omega_{2}}J_{0}\right]|\Psi'_1\rangle=(E^2-m^2c^4)|\Psi'_1\rangle.\
\end{equation}
The exact energy spectrum for $g$ and $f$ defined as in equation (\ref{part}) is obtained from equation (\ref{spectrum})
\begin{equation}
E=\pm{mc^{2}}\sqrt{\hbar\left(2(\omega_{1}+\omega_{2})j+\sqrt{(\omega_{1}-\omega_{2})^{2}+4\omega_{1}\omega_{2}}\mu\right)+1},\label{jcspec}
\end{equation}
and the corresponding eigenfunctions are given by equation (\ref{estado}).

By writing $E=mc^2+\epsilon$, we obtain that the non-relativistic limit $(\epsilon\ll mc^2)$ of the uncoupled equation (\ref{pnu1}) becomes
\begin{equation}
\hbar\left[\omega_{1}a^{\dag}a+\omega_{2}b^{\dag}b+\sqrt{\omega_{1}\omega_{2}}(b^{\dag}a+a^{\dag}b)\right]|\Psi_1\rangle.
\end{equation}
Formally, this is the Hamiltonian of the time-independent problem of two coupled oscillators \cite{Loui,Fan}, where we have identified the term $\sqrt{\omega_{1}\omega_{2}}$ with the coupling constant $\lambda$ of the oscillators. Moreover, the relativistic energy spectrum of equation (\ref{jcspec}), in the non-relativistic limit is in full agreement with the energy spectrum of this problem \cite{Nos2}. Therefore, the generalized Jaynes-Cummings model of equation (\ref{GJC}) can be considered a relativistic version of the two coupled oscillators, with a particular choice of the complex parameters $f$ and $g$.

\section{Concluding remarks}

We introduced two generalizations of the Jaynes-Cummings and Anti-Jaynes-Cummings models with two modes of oscillation. In the first generalization, we considered a linear combination of the JC and AJC models and showed that this system possesses the $SU(1,1)$ symmetry. By using the Jordan-Schwinger realization of this algebra, the tilting transformation and the $SU(1,1)$ number coherent states of the two-dimensional harmonic oscillator, we obtained the energy spectrum and eigenfunctions of this new model. As a physical application we connected this generalized JC-AJC model, in the non-relativistic limit, with the non-degenerate parametric amplifier.

For the second generalization, we considered a liner combination of two JC models and showed that this system possesses the $SU(2)$ symmetry. By using the appropriate Jordan-Schwinger realization and the tilting transformation, we were able to obtain the energy spectrum. Also, it has been shown that the eigenfunctions of this model are the $SU(2)$ number coherent states of the two-dimensional harmonic oscillator. In the non-relativistic limit, we connected this model with the problem of two coupled oscillators.

The relevance of the JC-AJC model studied in this work has been proven, since this model was used to describe the $2+1$ Dirac oscillator (with frequency $\omega$) in the presence of a constant magnetic field $B$ with vector potential $A=\frac{B}{2}[x,-y,0]$ \cite{Chiral}.

\section*{Acknowledgments}
This work was partially supported by SNI-M\'exico, COFAA-IPN, EDI-IPN, EDD-IPN, SIP-IPN project number $20170632$.

\section{Appendix.}
\subsection{ $SU(1,1)$ Perelomov number coherent states}

Three operators $K_{\pm}, K_0$ close the $su(1,1)$ Lie algebra if they satisfy the commutation relations \cite{Vourdas}
\begin{eqnarray}
[K_{0},K_{\pm}]=\pm K_{\pm},\quad\quad [K_{-},K_{+}]=2K_{0}.\label{com}
\end{eqnarray}
The Casimir operator $K^{2}$ for any irreducible representation of this group is given by
\begin{equation}
K^2=K^2_0-\frac{1}{2}\left(K_+K_-+K_-K_+\right).
\end{equation}
The action of these four operators on the Fock space states
$\{|k,n\rangle, n=0,1,2,...\}$ is
\begin{equation}
K_{+}|k,n\rangle=\sqrt{(n+1)(2k+n)}|k,n+1\rangle,\label{k+n}
\end{equation}
\begin{equation}
K_{-}|k,n\rangle=\sqrt{n(2k+n-1)}|k,n-1\rangle,\label{k-n}
\end{equation}
\begin{equation}
K_{0}|k,n\rangle=(k+n)|k,n\rangle,\label{k0n}
\end{equation}
\begin{equation}
K^2|k,n\rangle=k(k-1)|k,n\rangle.
\end{equation}
Thus, a representation of the $su(1,1)$ algebra is determined by the number $k$, called the Bargmann index. The discrete series
are those for which $k>0$.

The displacement operator $D(\xi)$ is defined in terms of the creation and annihilation operators $K_+, K_-$ as
\begin{equation}
D(\xi)=\exp(\xi K_{+}-\xi^{*}K_{-}),\label{do}
\end{equation}
where $\xi=-\frac{1}{2}\tau e^{-i\varphi}$, $-\infty<\tau<\infty$ and $0\leq\varphi\leq2\pi$.
The so-called normal form of the displacement operator is given by
\begin{equation}
D(\xi)=\exp(\zeta K_{+})\exp(\eta K_{0})\exp(-\zeta^*K_{-})\label{normal},
\end{equation}
where  $\zeta=-\tanh(\frac{1}{2}\tau)e^{-i\varphi}$ and $\eta=-2\ln \cosh|\xi|=\ln(1-|\zeta|^2)$ \cite{Gerry}.

The $SU(1,1)$ Perelomov coherent states are defined as the action of the displacement operator $D(\xi)$
onto the lowest normalized state $|k,0\rangle$ as \cite{Perellibro}
\begin{equation}
|\zeta\rangle=D(\xi)|k,0\rangle=(1-|\zeta|^2)^k\sum_{n=0}^\infty\sqrt{\frac{\Gamma(n+2k)}{n!\Gamma(2k)}}\zeta^n|k,n\rangle.\label{PCN}
\end{equation}
The $SU(1,1)$ Perelomov number coherent state $|\zeta,k,n\rangle$ is defined as the action of the displacement operator $D(\xi)$ onto an arbitrary
excited state $|k,n\rangle$ \cite{Nos1,Nos3}
\begin{eqnarray}
|\zeta,k,n\rangle &=&\sum_{s=0}^\infty\frac{\zeta^s}{s!}\sum_{j=0}^n\frac{(-\zeta^*)^j}{j!}e^{\eta(k+n-j)}
\frac{\sqrt{\Gamma(2k+n)\Gamma(2k+n-j+s)}}{\Gamma(2k+n-j)}\nonumber\\
&&\times\frac{\sqrt{\Gamma(n+1)\Gamma(n-j+s+1)}}{\Gamma(n-j+1)}|k,n-j+s\rangle.\label{PNCS}
\end{eqnarray}

The similarity transformations $D^{\dag}(\xi)K_{+}D(\xi)$, $D^{\dag}(\xi)K_{-}D(\xi)$, and
$D^{\dag}(\xi)K_{0}D(\xi)$ of the $su(1,1)$ Lie algebra generators are computed by using the displacement operator $D(\xi)$ an the Baker-Campbell-Hausdorff identity
\begin{equation*}
e^{A}Be^{-A}=B+[B,A]+\frac{1}{2!}[[B,A],A]+\frac{1}{3!}[[[B,A]A]A]+...
\end{equation*}
These results explicitly are
\begin{equation}
D^{\dag}(\xi)K_{+}D(\xi)=\frac{\xi^{*}}{|\xi|}\alpha K_{0}+\beta\left(K_{+}+\frac{\xi^{*}}{\xi}K_{-}\right)+K_{+},\label{DK+}
\end{equation}
\begin{equation}
D^{\dag}(\xi)K_{-}D(\xi)=\frac{\xi}{|\xi|}\alpha K_{0}+\beta\left(K_{-}+\frac{\xi}{\xi^{*}}K_{+}\right)+K_{-},\label{DK-}
\end{equation}
\begin{equation}
D^{\dag}(\xi)K_{0}D(\xi)=(2\beta+1)K_{0}+\frac{\alpha\xi}{2|\xi|}K_{+}+\frac{\alpha\xi^{*}}{2|\xi|}K_{-},\label{DK0}
\end{equation}
where $\alpha=\sinh(2|\xi|)$ and $\beta=\frac{1}{2}\left[\cosh(2|\xi|)-1\right]$.

A particular realization of the $su(1,1)$ Lie algebra is given by the Jordan-Schwinger operators
\begin{equation}
K_0=\frac{1}{2}\left(a^{\dag}a+b^{\dag}b+1\right), \quad K_+=a^{\dag}b^{\dag},\  \quad K_-= ba,\label{gen11}
\end{equation}
where the two sets of operators $(a,a^{\dag})$ and $(b,b^{\dag})$ satisfy the bosonic algebra
\begin{equation}
[a,a^{\dag}]=[b,b^{\dag}]=1, \quad\quad[a,b^{\dag}]=[a,b]=0.
\end{equation}
If $N_d$ is the difference of the number operators of the two oscillators, then $N_d$ commutes with all
the generators of this algebra and the Casimir operator is given by \cite{vourdasanalytic}
\begin{equation}
K^2=\frac{1}{4}N_d^2-\frac{1}{4}, \quad\quad N_d=b^{\dag}b-a^{\dag}a,\nonumber
\end{equation}
\begin{equation}
[N_d,K_0]=[N_d,K_+]=[N_d,K_-]=0.
\end{equation}

\subsection{ $SU(2)$ Perelomov number coherent states}

The $su(2)$ Lie algebra is spanned by the generators $J_{+}$, $J_{-}$ and $J_{0}$, which satisfy the commutation relations \cite{Vourdas}
\begin{eqnarray}
[J_{0},J_{\pm}]=\pm J_{\pm},\quad\quad [J_{+},J_{-}]=2J_{0}.\label{com2}
\end{eqnarray}
The Casimir operator $J^2$ in this representation is defined as
\begin{equation}
J^{2}=J_0^2+\frac{1}{2}\left(J_+J_-+J_-J_+\right).
\end{equation}
The action of these four operators on the Dicke space states (angular momentum states)\\
$\{|j,\mu\rangle, -j\leq\mu\leq j\}$ is
\begin{equation}
J_{+}|j,\mu \rangle=\sqrt{(j-\mu)(j+\mu+1)}|j,\mu+1 \rangle,\label{j+m}
\end{equation}
\begin{equation}
J_{-}|j,\mu \rangle=\sqrt{(j+\mu)(j-\mu+1)}|j,\mu-1 \rangle,\label{j-m}
\end{equation}
\begin{equation}
J_{0}|j,\mu \rangle=\mu|j,\mu \rangle,\label{j0m}
\end{equation}
\begin{equation}
J^2|j,\mu \rangle=j(j+1)|j,\mu \rangle.
\end{equation}
The displacement operator $D(\xi)$ for this Lie algebra is given by
\begin{equation}
D(\xi)=\exp(\xi J_{+}-\xi^{*}J_{-}),\label{D}
\end{equation}
where $\xi=-\frac{1}{2}\theta e^{-i\varphi}$. By means of Gaussian decomposition we can obtain the normal form of this operator
\begin{equation}
D(\xi)=\exp(\zeta J_{+})\exp(\eta J_{0})\exp(-\zeta^*J_{-})\label{normal2},
\end{equation}
where $\zeta=-\tanh(\frac{1}{2}\theta)e^{-i\varphi}$ and $\eta=-2\ln \cosh|\xi|=\ln(1-|\zeta|^2)$.

The $SU(2)$ Perelomov coherent states $|\zeta\rangle=D(\xi)|j,-j\rangle$ are defined as  \cite{Perellibro,Arecchi}
\begin{equation}
|\zeta\rangle=\sum_{\mu=-j}^{j}\sqrt{\frac{(2j)!}{(j+\mu)!(j-\mu)!}}(1+|\zeta|^{2})^{-j}\zeta^{j+\mu}|j,\mu\rangle.\label{PCN2}
\end{equation}
Hence, we can define the $SU(2)$ Perelomov number coherent states $|\zeta,j,\mu\rangle$ as the action of the displacement operator $D(\xi)$ onto an arbitrary
excited state $|j,\mu\rangle$ \cite{Nos2,Nos3}
\begin{eqnarray}
|\zeta,j,\mu\rangle &=&\sum_{s=0}^{j-\mu+n}\frac{\zeta^{s}}{s!}\sum_{n=0}^{\mu+j}\frac{(-\zeta^*)^{n}}{n!}e^{\eta(\mu-n)}
\frac{\Gamma(j-\mu+n+1)}{\Gamma(j+\mu-n+1)}\nonumber\\ &&\times\left[\frac{\Gamma(j+\mu+1)\Gamma(j+\mu-n+s+1)}{\Gamma(j-\mu+1)\Gamma(j-\mu+n-s+1)}\right]^{\frac{1}{2}}|j,\mu-n+s\rangle.\label{PNCS2}
\end{eqnarray}

The similarity transformations $D^{\dag}(\xi)J_{+}D(\xi)$, $D^{\dag}(\xi)J_{-}D(\xi)$, and $D^{\dag}(\xi)J_{0}D(\xi)$ of the $su(2)$ Lie algebra generators are computed by using the displacement operator $D(\xi)$ and the Baker-Campbell-Hausdorff identity  
\begin{equation}
D^{\dag}(\xi)J_{+}D(\xi)=-\frac{\xi^{*}}{|\xi|}\delta J_{0}+\epsilon\left(J_{+}+\frac{\xi^{*}}{\xi}J_{-}\right)+J_{+},
\end{equation}
\begin{equation}
D^{\dag}(\xi)J_{-}D(\xi)=-\frac{\xi}{|\xi|}\delta J_{0}+\epsilon\left(J_{-}+\frac{\xi}{\xi^{*}}J_{+}\right)+J_{-},
\end{equation}
\begin{equation}
D^{\dag}(\xi)J_{0}D(\xi)=(2\epsilon+1)J_{0}+\frac{\delta\xi}{2|\xi|}J_{+}+\frac{\delta\xi^{*}}{2|\xi|}J_{-},
\end{equation}
where $\delta=\sin(2|\xi|)$ and $\epsilon=\frac{1}{2}\left[\cos(2|\xi|)-1\right]$.

The Jordan-Schwinger realization of the $su(2)$ algebra is
\begin{equation}
J_0=\frac{1}{2}\left(a^{\dag}a-b^{\dag}b\right), \quad J_+=a^{\dag}b, \quad J_-=b^{\dag}a,\label{gen}
\end{equation}
where again, the two sets of operators $(a, a^{\dag})$ and $(b, b^{\dag})$ satisfy the bosonic algebra.
It is important to note that, besides the Casimir operator, there is another operator $N_s$ (called the number operator)
which commutes with all the generators of the $su(2)$ algebra. The Casimir operator $J^2$ for this realization 
is written in terms of this operator $N_s$ and is given by \cite{vourdasanalytic}
\begin{equation}
J^2=\frac{N_s}{2}\left(\frac{N_s}{2}+1\right),\quad\quad N_s=a^{\dag}a+b^{\dag}b,\nonumber
\end{equation}
\begin{equation}
[N_s,J_+]=[N_s,J_-]=[N_s,J_z]=0.
\end{equation}

\end{document}